\documentclass[floats,floatfix,showpacs,amssymb,prd,twocolumn,superscriptaddress,nofootinbib,nolongbibliography,reprint]{revtex4-2}

\usepackage{amssymb,amsmath,verbatim,mathtools,needspace,enumitem,etoolbox,graphicx,physics,microtype,afterpage,bigints,gensymb,tabularx,mathtools,siunitx, scalerel}
\usepackage[dvipsnames, usenames]{xcolor}
\definecolor{linkcolor}{rgb}{0.0,0.3,0.5}
\definecolor{dodgerblue}{HTML}{1E90FF}
\usepackage{hyperref}
\hypersetup{unicode, colorlinks=true, linkcolor=linkcolor, citecolor=linkcolor, filecolor=linkcolor,urlcolor=linkcolor, pdfusetitle}
\usepackage[all]{hypcap}
\usepackage[T1]{fontenc}
\usepackage[utf8]{inputenc}
\usepackage{orcidlink}
\usepackage [english]{babel}
\usepackage [autostyle, english = american]{csquotes}
\MakeOuterQuote{"}
\usepackage{soul}
\usepackage{bbm}

\renewcommand{\emph}[1]{\textit{#1}}

\newcommand{\umani}{\affiliation{Department of Physics and Astronomy \& Winnipeg Institute for Theoretical Physics, University of Manitoba, Winnipeg, R3T 2N2, Canada}}

\newcommand{\nottingham}{\affiliation{Nottingham Centre of Gravity \& School of Mathematical Sciences, University of Nottingham, University Park, Nottingham, NG7 2RD, United Kingdom}}

\newcommand{\paris}{\affiliation{Département de Mathématiques, ENSTA Paris, 91120 Palaiseau, France}}

\newcommand{\Msol}{\rm \,M_{\odot}}

\begin{document}

\title{Simulation-calibrated Bayesian inference for progenitor properties of the microquasar SS\,433}

\author{Nathan Steinle$\,$\orcidlink{0000-0003-0658-402X}}
\email{nathan.steinle@umanitoba.ca}
\umani

\author{Matthew Mould$\,$\orcidlink{0000-0001-5460-2910}}
\nottingham

\author{Sarah Al Humaikani$\,$}
\paris

\author{Austin MacMaster$\,$}
\umani

\author{Brydyn Mac Intyre$\,$}
\umani

\author{Samar Safi-Harb$\,$\orcidlink{0000-0001-6189-7665}}
\umani

\begin{abstract} 
SS\,433 is one of the most extreme Galactic X-ray binaries, launching semi-relativistic jets and showing clear signs of super-critical accretion onto what is likely a black hole. Yet the properties of the binary system that produced it remain uncertain. To solve the inverse problem of inferring the progenitor properties of binaries that evolve into SS\,433-like systems, we use an iterative, simulation-based calibration framework that combines Bayesian inference with the isolated binary-evolution code \textsc{COSMIC}. Using six measured properties of SS\,433 and the dynamic nested sampler \texttt{dynesty}, we explore a ten-dimensional space of possible progenitor masses, orbits, mass-transfer histories, and natal-kick velocities. This approach identifies the regions of parameter space most consistent with SS\,433 and allows us to iteratively refine the resulting progenitor distributions. We find 90\% confidence intervals for the progenitor initial primary mass of $(8, 11),\mathrm{M_\odot}$, secondary mass of $(32, 40),\mathrm{M_\odot}$, orbital period of $(136, 2259)$ days, eccentricity of $(0.26, 0.6)$, common-envelope efficiency of $(0.44, 0.76)$, accreted fraction during stable mass transfer of $(0.22, 0.6)$, and black-hole natal-kick magnitude of $(5, 68)$ km/s. These results show that direct probabilistic inference of X-ray binary progenitors can yield new constraints on the formation of extreme accretion systems like SS\,433, which has important implications for theoretical expectations of the population of SS\,433-like systems in the Galaxy and their connection with cosmic ray observations.
\end{abstract}

\maketitle

\section{Introduction} 
\label{sec:Intro}

Stellar-mass black holes (BHs) in binary systems are typically identified through observations of the electromagnetic emission from an accretion disk fueled by mass transfer from a stellar companion   \citep{2014SSRv..183..223C,2016A&A...587A..61C}. When the companion is a high-mass star (i.e., $m_* \gtrsim 8$~$\Msol$), the system is classified as a high-mass X-ray binary (HMXB) \cite{2023A&A...677A.134N,2023A&A...671A.149F}. Spectroscopy of the donor star can provide its luminosity, and parameters such as BH mass, orbital eccentricity, and period can be constrained through photometry, astrometry, and radial velocities via the binary mass function \citep{2022NatAs...6.1085S,2023MNRAS.521.4323E,2023MNRAS.518.1057E}. These objects are also typically associated with astrophysical jets, offering further information on their evolution. 

The diversity of detected HMXBs, including both X-ray-bright and X-ray quiescent systems, offers snapshots of different stages in binary evolution. Modeling these systems has been a main driver in the development of binary population synthesis tools, such as \textsc{BSE} \cite{2002MNRAS.329..897H}, \textsc{COSMIC} \citep{2020ApJ...898...71B}, \textsc{StarTrack} \citep{2008ApJS..174..223B}, \textsc{COMPAS} \citep{2022ApJS..258...34R}, \textsc{BPASS} \citep{2017PASA...34...58E}, and \textsc{POSYDON} \citep{2023ApJS..264...45F}, and many others. 
Such models allow one to investigate astrophysical processes, which are abundant and uncertain, and theoretical evolutionary pathways. 

SS\,433 is a rare and valuable case—a persistent microquasar embedded in the W50 nebula and characterized by its precessing, semi-relativistic jets, high mass-transfer rate, and likely BH accretor \citep{2008ApJ...676L..37H,2011MNRAS.417.2401B}. Its extreme properties and rich archival data make it an ideal testing ground for data-driven modeling. 
As several of the binary parameters have been observationally constrained, the aim of our study is to leverage this knowledge to explore the properties of plausible progenitor binaries of SS\,433. 

Traditionally, estimation of the progenitor properties of X-ray binaries like SS\,433 relies on dense grid-based sampling \cite{2020ApJ...896...34H}, which can be prohibitively expensive in high dimensions. Recent years have seen a growing adoption of Bayesian and machine learning methods to constrain the properties of observed compact-object binaries. 
For instance, Ref.~\cite{2025PhRvD.111j3053S} used a conditional variational autoencoder to accelerate parameter estimation by constraining the prior range used in the usual BH binary parameter estimation methodology. 
Ref.~\cite{2020PhRvL.124d1102C} present a neural network inference framework for rapid approximation of gravitational wave event posteriors. 
Simulation-based inference has been used for parameter estimation of intermediate-mass black hole binaries \cite{2024arXiv240603935R}, and combined with normalizing flows for bypassing the `global fit' problem, i.e., the search for single signals among many overlapping gravitational-waves, of the future LISA detector \cite{2025arXiv250622543S}.  

More specifically, such methods have found particular utility in estimating the possible progenitors of observed compact binaries, 
often solving a Bayesian inversion problem by performing inference on the progenitor posteriors from observed compact binary evidence.   
These approaches typically combine forward modeling, e.g., binary population synthesis, together with probabilistic inference to estimate initial binary parameters. 
For example, Ref.~\cite{2018ApJS..237....1A,2021ApJ...914L..32A} couple binary population synthesis with Markov Chain Monte Carlo (MCMC) sampling to infer progenitor parameters consistent with observed X-ray binaries and gravitational-wave sources, and  Ref.~\cite{2023ApJ...950..181W} implemented a hierarchical Bayesian model combining observed gravitational-wave events with population synthesis models to reconstruct the distribution of massive binary progenitors. 
Recent works applied normalizing flows to emulate expensive binary population models and enable fast probabilistic inference of progenitor distributions \cite{2025arXiv250303819C,2025arXiv250605657R}, 
polynomial chaos expansion and Bayesian MCMC to efficiently recover progenitor posteriors \cite{2025arXiv250600757S}, and neural variational learning to accelerate binary BH population inference \cite{2025PhRvD.111l3049M}. 

Our methodology utilizes similar principles as these approaches: we use \texttt{dynesty} \cite{2020MNRAS.493.3132S} to sample a ten-dimensional zero-age main sequence (ZAMS) progenitor space represented by a multivariate Gaussian likelihood constructed from the observed HMXB properties of SS\,433 and constrained by \textsc{COSMIC} \citep{2020ApJ...898...71B} binary evolution. 
Our approach is an example of inverse (or reverse) Bayesian modeling of binary progenitor parameters, guided by principles of Bayesian optimization to efficiently explore high-likelihood regions without surrogate emulation (similar to the method of \cite{2023ApJ...950..181W}), enabling statistical inference of the progenitor properties of SS\,433. 
We find that uncalibrated Bayesian sampling struggles to produce posterior samples corresponding to SS\,433-like systems with high confidence. We solve this issue by applying simulation-based calibration, which allows us to iteratively build reliable posterior distributions of ZAMS parameters. 

This paper is organized as follows. We present our Bayesian framework in Sec.~\ref{sec:Meth}, showcase our results in Sec.~\ref{sec:Results}, and summarize our conclusions and discussions in Sec.~\ref{sec:Conclusions}. 

\section{Methods}
\label{sec:Meth}

\subsection{The source: \emph{SS\,433}}

SS\,433 is a unique Galactic microquasar \cite{1999ARA&A..37..409M} and a high-mass X‑ray binary associated with the W50 nebula, commonly believed to be a supernova remnant \cite{2004ASPRv..12....1F}. The system comprises a compact object—strongly evidenced to be a $\approx$ 5--15$\Msol$ stellar BH—accreting material from a likely A‑type ($\approx$10–24$\Msol$) supergiant companion via a supercritical accretion disk \cite{2004ASPRv..12....1F}. SS\,433 exhibits canonical bipolar relativistic jets moving at $\approx$0.26c that precess with a $\approx$162 day period \cite{1979MNRAS.187P..13F,1979Natur.279..701A}, giving rise to moving emission lines whose Doppler shifts were first observed in the late 1970s \cite{1984ARA&A..22..507M,1979Natur.279..701A}. The binary orbit has a period of $\approx$13.1 days \cite{1981ApJ...251..604C} and displays eclipses due to an inclination around 79$^{\circ}$ \cite{2008ARep...52..487D}, enabling precise orbital studies. The discovery of persistent moving optical lines led to the classification of SS\,433 as the Milky Way’s first known microquasar \cite{2004ASPRv..12....1F}.

Extensive multi-wavelength observations across the electromagnetic spectrum---from radio/infrared to X-rays and TeV $\gamma$‑rays---have revealed precious details of SS\,433. Radio and optical spectroscopy have mapped the precessing jets and inferred their speed and geometry \cite{1993A&A...270..177V,1993A&A...270..204V,2004MNRAS.354.1239S,2004ApJ...616L.159B}, and coupled Chandra and VLA data revealed the structure of the arcsecond-scale jets \cite{2008ApJ...682.1141M,2013ApJ...775...75M}. 
Various Chandra and XMM‑Newton observations have resolved extended X‑ray jets and detected an accretion‑disk corona \cite{2002ApJ...564..941M,2005MNRAS.358..860M,2005A&A...431..575B}. 
INTEGRAL and hard X‑ray surveys (18–60 keV) have modeled eclipses and precessional variability, differentiating between narrow jet and extended corona emission regions \cite{2003A&A...411L.441C,2010MNRAS.402..479M,2010PASJ...62..323K}. In TeV $\gamma$‑rays, HAWC and H.E.S.S. recently detected and resolved high-energy gamma-ray emission \cite{2018Natur.562...82A, 2024Sci...383..402H}
spatially aligned with the X-ray bright, non-thermal emitting lobes in W50 \cite{SafiHarb2022ApJ935}, indicating electron acceleration up to hundreds of TeV within W50’s lobes and confirming W50-SS\,433 as a promising Galactic PeVatron \cite{1999ApJ...512..784S, SafiHarb2022ApJ935, 2018Natur.562...82A, 2024Sci...383..402H}. LHAASO has reported ultra-high-energy (i.e., $> 100$ TeV) $\gamma$-ray emission associated with SS\,433, spatially coincident with a giant atomic/molecular cloud near W50, and demonstrated that the emission cannot be described by a single leptonic component, providing compelling evidence that hadronic processes in SS\,433 and W50 accelerate particles to energies well above the PeV scale \cite{2024arXiv241008988L}.

Decades of combined observational datasets have tightened constraints on binary parameters. Long-term photometry uncovered a slight orbital eccentricity of $e = 0.05 \pm 0.01$ \cite{2021MNRAS.507L..19C}. 
Estimates of the binary component masses can vary depending on the datasets used and modeling assumptions employed. 
The mass of the accreting compact object is  constrained to be $M_X \gtrsim 5 \Msol$ indicating a black hole \cite{2004ApJ...615..422H,2008ApJ...676L..37H,2018MNRAS.479.4844C,2019MNRAS.485.2638C,2018A&A...619L...4B,2023NewA..10302060C}, i.e., a binary mass ratio $q \gtrsim$ 0.6, with $M_X \approx 5$--$9 \Msol$ and donor mass $\sim$8–15$\Msol$ \cite{2019MNRAS.485.2638C}; however, other observations cast this into doubt, e.g., ~\cite{2010ApJ...709.1374K}.  
Hydrodynamic simulations coupling precessing jets with the W50 shell reproduce its elongated morphology and estimate the jet interactions and nebula age which is constrained to be $\sim \num{2e4}$ yrs  \cite{2011MNRAS.414.2838G,2021ApJ...910..149O}.  
These measurements reveal SS\,433’s orbital dynamics and evolutionary stage as a binary with a black hole accreting from its stellar companion. The relevant measurements for our study are compiled in Table~\ref{Tab:Measurements}. 

The progenitor of SS\,433 likely began as a high‑mass binary which underwent a supernova forming a black hole. Population synthesis modeling \cite{2020ApJ...896...34H} suggests SS\,433 entered Roche‑lobe overflow during the donor’s Hertzsprung-gap phase, with current donor mass $\approx$ 24$\Msol$ and $M_X \approx 8 \Msol$, consistent with observed mass and age constraints. 
The discovery of an X-ray shell north of SS\,433 and coincident with a polarized radio shell was used to support the supernova remnant shell interpretation of the W50 nebula and infer the nature of its progenitor star \cite{2024ApJ...975L..28C}. When coupled with the results of hydrodynamical supernova simulations, this suggests that the mass of the progenitor star that formed into a BH was $\gtrsim 25 \Msol$. 
Theoretical work also demonstrates that systems with $q \gtrsim$ 0.7 can avoid entering a common‑envelope phase and instead evolve in a stable, semi‑detached configuration via isotropic re‑emission—matching SS\,433’s behavior \cite{2017MNRAS.471.4256V}.
As we shall see in the next subsection, these evolutionary possibilities are generically encapsulated by binary population models. 

\newcolumntype{C}[1]{>{\centering\arraybackslash}p{#1}}

\begin{table*}
\caption{
Observations of the parameters of the X-ray binary SS433, the compact object mass $m_{\rm X}$, the stellar companion mass $m_*$, the orbital period $P$ and eccentricity $e$, the mass transfer rate $\dot{M}$, and the estimate of the age $t_{\rm SNR}$ of the supernova remnant W50 which we will use as the age of the compact object in SS433. Note that not all parameters are provided in each study. A subset of these parameter measurements are used to define the multivariate-Gaussian progenitor likelihood for our analysis. 
}
\label{Tab:Measurements}
\vspace{0.1cm}
\def\arraystretch{1.2}
\centering
\begin{tabular}{l||C{0.75in}|C{0.75in}|C{0.85in}|C{0.75in}|C{0.75in}|C{0.75in}}
  \hline
  \hline
  Study & $m_{\rm X}$ [$\Msol$] & $m_*$ [$\Msol$] & $P$ [days] & $e$ & $\dot{M}$ [$\Msol$/yr] & $t_{\rm SNR}$ [kyr]\\
  \hline
  Gies et al. (2002) \cite{2002ApJ...578L..67G} & $11\pm5$   & $19\pm7$  &  —   & —   & —  & — \\
  Hillwig et al. (2004, 2008) \cite{2004ApJ...615..422H, 2008ApJ...676L..37H} & $2.9\pm0.7$  & $10.9\pm3.1$  &  —   & —   & —  & — \\
  Cherepashchuk et al. (2018) 
  \cite{2018MNRAS.479.4844C}& $>7$    & $>12$   & —  & —   & — & — \\
  Cherepashchuk et al. (2019) \cite{2019MNRAS.485.2638C} &   $>5\text{–}9$  & $8\text{–}15$ & —  &  —    & —  & — \\
  Crampton \& Hutchings  (1981) \cite{1981ApJ...251..604C} \footnote{Long‑term followup}  & — & — & 13.082 $\pm 0.003$   & —   & — & — \\
  Cherepashchuk et al. (2021) \cite{2021MNRAS.507L..19C} \footnote{Long‑term photometry 1979–2020}   & — & —  &  —   & $0.05\pm0.01$   & —    & — \\
  King et al. (2000) \cite{2000ApJ...530L..25K}  
  \footnote{Other estimates are similar, e.g.'s  \cite{2004ASPRv..12....1F,2009MNRAS.397..849P}} & — & — & — & — & $\sim 10^{-4}$ & — \\
  Goodall et al. (2011) \cite{2011MNRAS.414.2838G,2021ApJ...910..149O} \footnote{Hydro‑simulations of jet–SNR W50 interaction} & — & — & — & — & — &  $17\text{–}21$ \\
  Chi et al. (2024) \cite{2024ApJ...975L..28C, 1998AJ....116.1842D} & — & — & — & — & — &  $20\text{–}30$ \\
  \hline
  \hline
\end{tabular}
\end{table*}

\subsection{Binary population synthesis}
\label{Subsec:Popsynth}

Stellar-mass black holes (BHs), i.e., with masses $m \lesssim 100 \Msol$, have been found in X-ray binaries from electromagnetic observations and with typical mass $m \sim 15$ M$_{\odot}$ \cite{2012ApJ...757...36K}. 
These binaries are generally thought to be in an intermediate stage of stellar binary evolution where one star has already formed a BH from gravitational core-collapse and the companion star has overflowed its Roche lobe causing mass transfer episodes. The nature of such mass transfer is sensitive to the stage of evolution of the donor star, the binary orbital properties, and the masses and radii of the binary components. 

Natal kicks are the main mechanism suspected to provide spin--orbit misalignments for binaries that evolve in isolation, however it is uncertain whether BHs receive recoil velocities. 
Historically, the population of X-ray binaries were thought to not receive natal kicks, however this is uncertain \cite{2012MNRAS.425.2799R,2013MNRAS.434.1355J}.
Before GW detections, X-ray binaries were the only available source for constraining kicks of BHs. 
This is especially interesting when one considers possible relationships between the BH natal kick and the precession of the jet in an XRB, where gravitational waves suggest the possibility for multi-messenger studies of XRB jets, which we leave to future work.  

In this study, we employ the rapid binary population synthesis code \textsc{COSMIC} for modeling the evolution and formation of HMXBs. 
We assume a solar-like metallicity $Z = 0.02$, and we use default parameter values in \textsc{COSMIC} unless otherwise specified. 
For detailed explanations of the various physical processes and models implemented in the \textsc{COSMIC} code, we refer the interested reader to, e.g., Ref.~\cite{2020ApJ...898...71B}.  
Here we will only briefly introduce the main processes relevant for our study: stable mass transfer, common envelope evolution, and BH natal kicks. 

Mass transfer processes dominate the evolution of a stellar binary's orbital angular momentum. As with other similar population synthesis models, \textsc{COSMIC} treats mass transfer in terms of time-dependent mass-radius exponents that relate the dynamical dependence of the donor and accretor radii and Roche lobe radii to the exchange of gas and that are sensitive to the stellar stage of the donor. 
While \textsc{COSMIC} takes into account a wide range of possible outcomes, two main possibilities are (dynamically) stable mass transfer (SMT) and common envelope evolution (CEE). 

In SMT, a fraction $f_{\rm a}$ of the donor's envelope is accreted, where the accretor's mass and spin angular momentum increase and the orbital separation can change under isotropic re-emission of the donor's envelope. The accretor can generally be any star or compact object, but here we focus on BH accretors. In population synthesis modeling, the accreted fraction is conventionally set to $f_{\rm a} = 0.5$ but its value has important implications for evolutionary pathways \cite{2022MNRAS.516.5737B}. 

In CEE, the mass transfer rate overwhelms the accretion rate and the donor's envelope subsumes the binary. The common envelope is expelled from the system at the expense of orbital gravitational energy which drastically decreases the orbital separation and possibly prematurely merges the binary. We use the $\alpha$ formalism for CEE with the free parameter $\alpha_{\rm CE}$, which encodes uncertainty in the binding energy of the star's envelope relative to changes in the gravitational orbital potential. 
A perfect binding energy corresponds to $\alpha_{\rm CE} = 1$, i.e., all orbital energy loss is used to expel the envelope, and values of $\alpha_{\rm CE}$ greater (lesser) than 1 correspond to binaries whose orbits shrink to larger (smaller) separations.

These two types of \emph{mass transfer events}, along with stellar winds mass-loss, are the main mechanisms by which the stars lose their envelopes and lead to possible formation of a compact object.  
Together, these possibilities sketch a simplified, yet complicated landscape of possible evolutionary pathways and their imprint on the correlations between the progenitor and observed binary properties. 

Similar methods are used to study the theoretical properties of source binaries of HMXBs and of gravitational-wave driven mergers.  
Currently, there are a handful of observed HMXBs \cite{2023hxga.book..143F} in the Milky Way and nearby satellites. 
Comparatively, the number of observed binary BH mergers has grown quickly over the last decade and is expected to accelerate exponentially in the coming years and in next decade with the advent of third-generation gravitational-wave detectors \cite{CE2021,ET2020,2024ApJ...969..108B}. 
The observed population of HMXBs are unlikely progenitors of detectable gravitational wave populations due to their differences in redshift and metallicity dependence \cite{2022ApJ...929L..26F,2022ApJ...938L..19G,2023MNRAS.524..245R}. 
However, this implies that our understanding of the formation of binaries and their progenitors may be revealed with greater certainty in the future by combining knowledge gained from both classes of sources. 

The complementarity between studies of HMXBs and gravitational wave mergers can be seen, for example, when considering the progenitors of a known binary. 
One can attempt to produce \textsc{COSMIC} output from grids of forward evolutions of binaries and select cuts from the output based on criteria specifying the properties of the desired HMXBs. 
This is the usual approach as demonstrated in Ref.~\cite{2020ApJ...896...34H}, who used a model for the jet of SS\,433 in conjunction with a binary population synthesis model to predict the likely XRB binary parameters of SS\,433 given observations of its jet. 

Instead, we utilize a machine-learning approach, based on an earlier method used to study the progenitors of gravitational-wave mergers and of X-ray quiet binaries \cite{2023ApJ...950..181W,2023MNRAS.518.1057E}, to statistically explore the progenitor binary star parameters informed by a set of XRB measurements of SS\,433, as detailed in the following subsections, providing a new look into the complexity of the XRB progenitor parameter space. 
We expect that the parameters that control mass transfer, i.e., SMT and CEE, and natal kicks are most important for determining the binary properties. 
There are many other parameters in a population synthesis framework like \textsc{COSMIC} that are also important for guiding the shape of the parameter space volume, highlighting higher order uncertainties, and our choice of progenitor parameters to sample is an approximation of the fuller parameter space.  

\subsection{Bayesian inference}

Our methodology belongs to the class of Bayesian approaches, where one infers latent input parameters of a physical system conditioned on observed outputs with a forward simulation model to evaluate their consistency. This class of methods is increasingly used across statistics and computational physics for solving inverse problems, i.e., those where the likelihood function may be costly to evaluate or analytically intractable.

In this work, we apply a Bayesian approach---similar to the method of Ref.~\cite{2023ApJ...950..181W} used for gravitational-wave sources---to the HMXB system SS\,433 by treating ZAMS binary parameters as latent inputs and evolving them forward using the \textsc{COSMIC} binary population synthesis model. We define a multivariate Gaussian likelihood over the six observed properties of SS\,433, corresponding to the six columns of Table~\ref{Tab:Measurements}, the black hole mass $m_{\rm BH}$, stellar donor mass $m_{\rm *}$, orbital period $P$, eccentricity $e$, mass transfer rate $\dot{\rm M}$, and black hole age $t_{\rm BH}$. 
We use nested sampling (via the \texttt{dynesty} sampler \cite{2020MNRAS.493.3132S}) to explore the posterior distribution of the following progenitor parameters: the initial masses $m_{1, \rm ZAMS},\ m_{2, \rm ZAMS}$, orbital period $P_{\rm ZAMS}$, eccentricity $e_{\rm ZAMS}$, common envelope efficiency $\alpha_{\rm CE}$, accreted fraction $f_{\rm acc}$, and natal kick velocity magnitude $v_{\rm k}$ and direction $\theta_{\rm k},\ \phi_{\rm k},\ \psi_{\rm k}$.
Additionally, our method incorporates an evolutionary feasibility constraint—i.e., whether the ZAMS binary evolves into a system resembling SS\,433—thus framing our approach as a constraint-aware global posterior sampler. 
This hybrid structure allows for efficient inference over a high-dimensional latent space, and offers a new path for reverse-engineering the progenitors of X-ray binaries using population-synthesis models. 

\subsubsection{Progenitor likelihood}

From the description of COSMIC binary evolution in the previous subsection, we have a forward model $f$ for the properties $y=f(x)$ of the observed source in terms of its progenitor properties $x$. 
We take $x = \{ m_{1, \rm ZAMS},\ m_{2, \rm ZAMS},\ P_{\rm ZAMS},\ e_{\rm ZAMS},\ \alpha_{\rm CE},\ f_{\rm acc},\ v_{\rm k},$ $\theta_{\rm k},\ \phi_{\rm k},\ \psi_{\rm k} \}$ for the initial binary parameters, and $y = \{ m_{\rm BH},\ m_{\rm *},\ P,\ e,\ \dot{\rm M},\ t_{\rm BH} \}$ for the HMXB parameters. 
Given a measurement of the source properties, we now wish to measure the progenitor properties.

Using Bayes' theorem with a chosen prior $\pi(x)$ on the progenitor properties, the progenitor posterior may be written as
\begin{align}
p(x|d)
\propto
\pi(x) p(d|x)
=
\pi(x) \int \dd{y} p(d|y) p(y|x)
\, .
\end{align}
The mapping from progenitor to observed binary is a deterministic function given by the population synthesis model $f$, i.e., $p(y|x) = \delta(y-f(x))$, where $\delta$ is the Dirac delta function. We model the likelihood function $p(d|y)$ as a multivariate Gaussian $\mathcal{G}(y|\mu,\sigma)$ in the observed properties of SS\,433 with diagonal covariance matrix, where the means $\mu$ and variances $\sigma^2$ are determined from observations listed in Table~\ref{Tab:Likelihood}. Altogether, the posterior is therefore
\begin{align}
p(x|d) \propto p(x) \mathcal{G}(y=f(x)|\mu,\sigma)
\, .
\label{Eq:SamplePosterior}
\end{align}

\newcolumntype{C}[1]{>{\centering\arraybackslash}p{#1}}

\begin{table}
\caption{
The means (top row) and variances (bottom row)  \footnote{\cite{1981ApJ...251..604C,2019MNRAS.485.2638C,2021MNRAS.507L..19C,2000ApJ...530L..25K,2011MNRAS.414.2838G}} used to construct the multivariate Gaussian likelihood of the observations of the six parameters of the SS433 system: the black hole mass $m_{\rm BH}$, the stellar companion mass $m_*$, the orbital period $P$ and eccentricity $e$, the mass transfer rate $\dot{M}$, and the age of the black hole $t_{\rm BH}$ which is the age of the supernova remnant W50. 
}
\label{Tab:Likelihood}
\vspace{0.1cm}
\def\arraystretch{1.2}
\centering
\begin{tabular}{C{0.5in}|C{0.5in}|C{0.45in}|C{0.45in}|C{0.55in}|C{0.5in}}
  \hline
  \hline
  $m_{\rm BH}$ [$\Msol$] & $m_*$ [$\Msol$] & $P$ [days] & $e$ & $\dot{M}$ [$\Msol$/yr] & $t_{\rm SNR}$ [kyr]\\
  \hline
    $7.0$ & $11.5$ &  $13.082$   &  $0.05$  &  $10^{-4}$  &  $19.0$  \\
    $0.6$ & $1.0$  &  $0.003$  &  $0.01$ &  $10^{-5}$  &  $0.001$ \\
  \hline
  \hline
\end{tabular}
\end{table}

\subsubsection{Optimized nested sampling}
\label{Subsec:Optimized}

We assume uniform prior distributions of the progenitor parameters, except for $P_{\rm ZAMS}$ and $\alpha_{\rm CE}$ whose priors are drawn from log-uniform distributions. The lower and upper bounds of the priors for the ten progenitor parameters are: 
$ m_{1,2 \rm ZAMS}\in[2,\ 100],\ 
P_{\rm ZAMS}\in[100,\ 10{,}000],\ 
e_{\rm ZAMS}\in[0,\ 1.0],\ 
\alpha_{\rm CE}\in[0.1,\ 20],\ 
f_{\rm acc}\in[0,\ 1],\ 
v_{\rm k}\in[0,\ 300],\ 
\cos\theta_{\rm k}\in[-1,\ 1],\ 
\phi_{\rm k}\in[0,\ 2\pi],\ 
\psi_{\rm k}\in[0,\ 2\pi]$. 
To sample from the joint posterior over the progenitor parameters defined in Eq.~(\ref{Eq:SamplePosterior}), we employ the dynamic nested sampler \texttt{dynesty} \cite{2020MNRAS.493.3132S}, utilizing the default settings and its MPI parallelization functionality. 

The likelihood definition in Eq.~(\ref{Eq:SamplePosterior}) alone is not sufficient to determine whether a binary is in an XRB phase. We enforce constraints on the evaluation of the likelihood to find binaries undergoing an XRB phase in a bound orbit. 
In this constraint function, we select binaries that have a BH for one component and that have just finished a phase of Roche-lobe overflow with a lower limit on the mass transfer rate of $10^{-9} \Msol/yr$, consistent with the assumption in Ref.~\cite{2020ApJ...896...34H}.  
For binaries that pass these checks, we obtain the six relevant parameters of the XRB binary and evaluate
Eq.~(\ref{Eq:SamplePosterior}). The likelihood of sources which do not follow this pathway is set to zero. 

\subsubsection{Simulation-based calibration}
\label{Subsec:SBC}

When running \texttt{dynesty} on the fiducial priors specified in the previous section, not all of the ZAMS posteriors correspond to XRB parameters near to the SS\,433 parameters used to define the likelihood. This is a common issue with inverse modeling,
especially in high-dimensional parameter spaces with nontrivial features such as our binary population synthesis model.
To address this issue, we employ simulation-based calibration (SBC) to our workflow, an iterative strategy to narrow the ZAMS prior ranges on region(s) that produce the most SS\,433-like binaries possible given the Bayesian framework. This depends on each parameter used to define the likelihood by differing amounts, i.e., some XRB parameters will be farther from their true values, necessitating a per-parameter evaluation approach. 

In our iterative calibration pipeline, we combine forward population synthesis with a resampling strategy augmented by machine learning
to refine ZAMS prior bounds toward producing SS\,433–like systems. The core of this process begins after an initial \texttt{dynesty} run with wide priors (e.g., the fiducial prior bounds above), from which we save all attempted evolutions and all the successful posterior samples. To analyze these samples, we employ both nonparametric and parametric density estimation methods. 

We first reconstruct the one-to-one mapping between the accepted ZAMS posterior samples and their corresponding COSMIC XRB outputs. Concretely, we build a cKDTree \cite{2020SciPy-NMeth} over all saved ZAMS draws from the forward sweep and, for each \texttt{dynesty} posterior sample, retrieve the nearest saved ZAMS to recover its matched XRB, yielding an aligned array of six-dimensional observables per posterior draw. This step is algorithmic rather than learned (no model fitting), and assumes Euclidean proximity is sufficient for correct matching (i.e., the saved prior grid is dense and parameters are on comparable scales). Establishing this mapping is what enables the subsequent SBC-style self-consistency checks between simulated outputs and inferential behavior, in line with the original simulation-based calibration framework that validates Bayesian computation by checking rank and coverage diagnostics under the generative model \cite{2018arXiv180406788T}.

Next, we use an adaptive scheme to select the target tightness of the SS\,433 acceptance window via a search over candidate values of the parameter $n_{\rm matches}$, which controls the minimum number of SS\,433-like binaries required after tightening. For each candidate, our routine: (i) adaptively tightens per-observable thresholds---a coordinate-descent with per-dimension binary searches finds the smallest per-parameter standard deviation from the values that define the likelihood (see Table~\ref{Tab:Likelihood}) which we combine into a $n\sigma$ vector whose joint rectangular window retains at least $n_{\rm matches}$ systems; (ii) model the resulting ZAMS subset with a Gaussian mixture model \cite{2011JMLR...12.2825P} (full covariance, $n_{\mathrm{components}} = 10$) to capture multimodality; (iii) resample from \textit{significant} components (default weight cutoff $1 / n_{\mathrm{components}}$, $n_{\mathrm{resample}} = 500$ per component); and (iv) convert the resampled cloud into proposed prior boxes using univariate kernel-density-estimation-based \cite{2020SciPy-NMeth} $1$--$99\%$ marginal quantiles. 
Each candidate’s proposal is scored to balance SS\,433-like recovery and parameter-space coverage: 
\begin{equation}\label{Eq:score}
    {\rm score} = p \frac{\log(n_{\mathrm{selected}})}{\log(V_{\rm ratio})}
    \, ,
\end{equation}
where $n_{\mathrm{selected}}$ is the number of found SS\,433 binaries, $p$ is the purity (i.e., fraction of sampled binaries within the new bounds and SS\,433-like) and $V_{\rm ratio} = V_{\rm new}/V_{\rm old}$ is the minimum ratio of prior volumes of the two \texttt{dynesty} runs allowable where the new and old prior volumes correspond to before and after an SBC iteration step, respectively. We assume conservative settings, $n_{\mathrm{selected}} \geq 5000$ and $V_{\mathrm{ratio}} > 10^{-7}$, to avoid over-learning and under-sampling. 
The candidate with the highest feasible score determines the optimal $n_{\rm matches}$ used in the final bound computation. 
In the case where an optimal $n_{\rm matches}$ is not found, we assume a conservative balance between $p$ and $V_{\mathrm{ratio}}$. 

With the chosen $n_{\rm matches}$ we obtain the new ZAMS prior bounds, clipping each dimension to its physical limits (e.g., $e \in [0,1]$, $\alpha_{\mathrm{CEE}} \in [0.1,20]$, $v_k \in [0,300]$). We then validate these bounds by first flagging all ZAMS posterior samples inside the proposed box and then measure the fraction whose forward-mapped XRBs fall within the per-parameter SS\,433 windows defined by the optimized $n\sigma$ vector (reporting counts, in-box fraction, and SS,433-like purity). This closing check serves as a posterior-conditional calibration diagnostic, assessing whether the inference remains well-calibrated in the region supported by the observed data—consistent with the ``posterior SBC'' perspective that extends classical SBC from prior-predictive to data-conditional validation \cite{2025arXiv250203279S}. 
With this machinery, we obtain the optimized $n\sigma$ vector and use it to compute the fraction $f_{\rm in}$ of binaries that are SS\,433-like and within the calibrated bounds, enabling us to build posterior distributions of SS\,433 progenitors more reliably than is possible with calibrated sampling. 

\section{Results}
\label{sec:Results}

\begin{figure*}
\centering
\includegraphics[width=\textwidth]{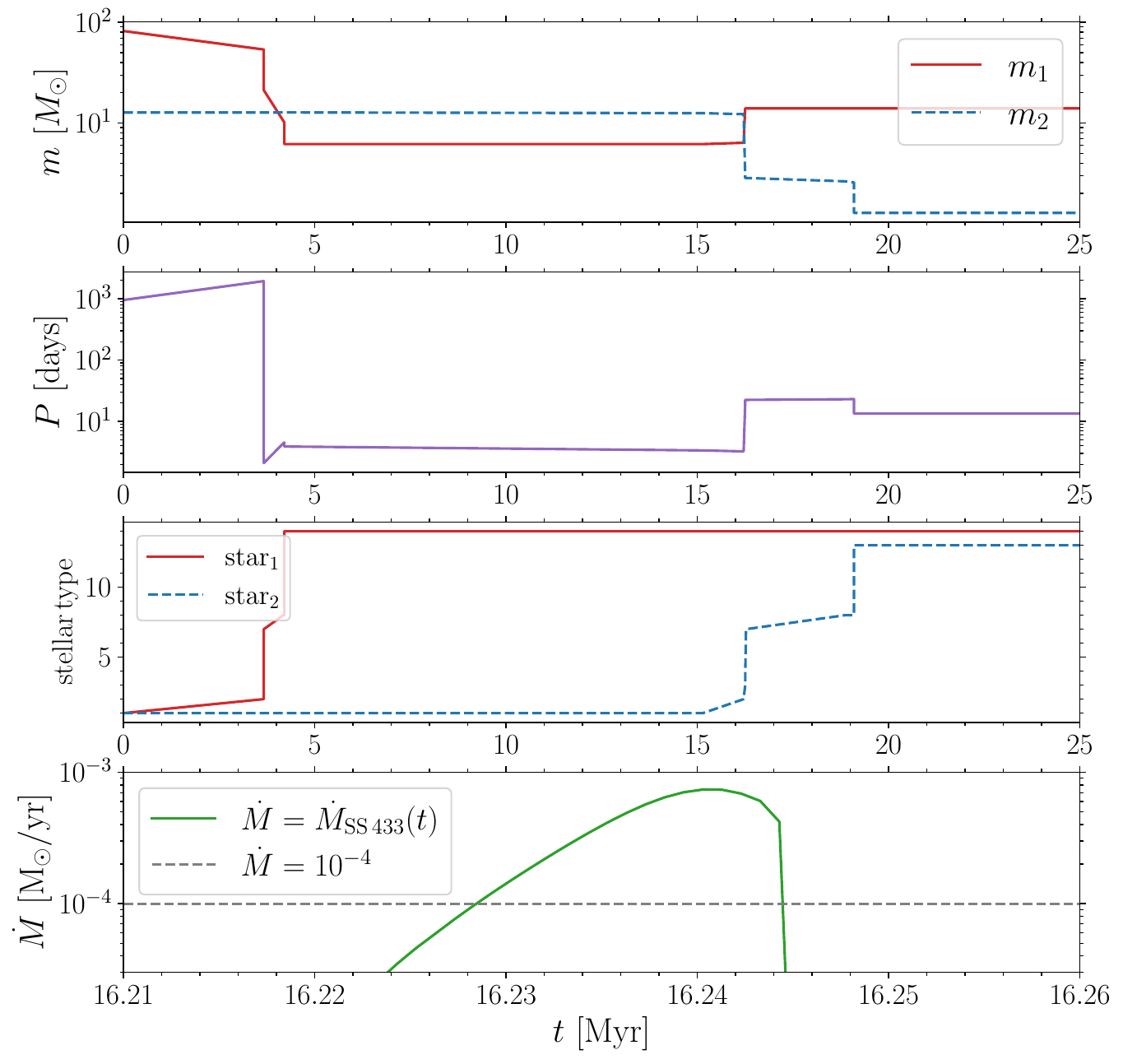}
\caption{The evolution of a single representative SS\,433-like binary generated from our inverse-Bayesian sampling of the \textsc{COSMIC} population synthesis model. The first three panels show the evolution of the binary masses, orbital period, and stellar types, respectively, over the lifetime of the binary. The properties of the initially more (less) massive star are shown with red solid (blue dashed) lines. The bottom panel shows a zoomed-in view of the phase of stable mass transfer where the initially more massive star that formed into a black hole accretes material from the initially less massive star; the corresponding mass transfer rate (green solid line) evolves in time and peaks at $\approx \num{7e-4} \Msol/yr$ and exceeds the observed mass transfer rate of SS\,433 (gray dashed line) of $10^{-4} \Msol/yr$ for about 0.01 Myr. 
} \label{F:SingleEvo}
\end{figure*}

\begin{figure*}
\centering
\includegraphics[width=\textwidth]{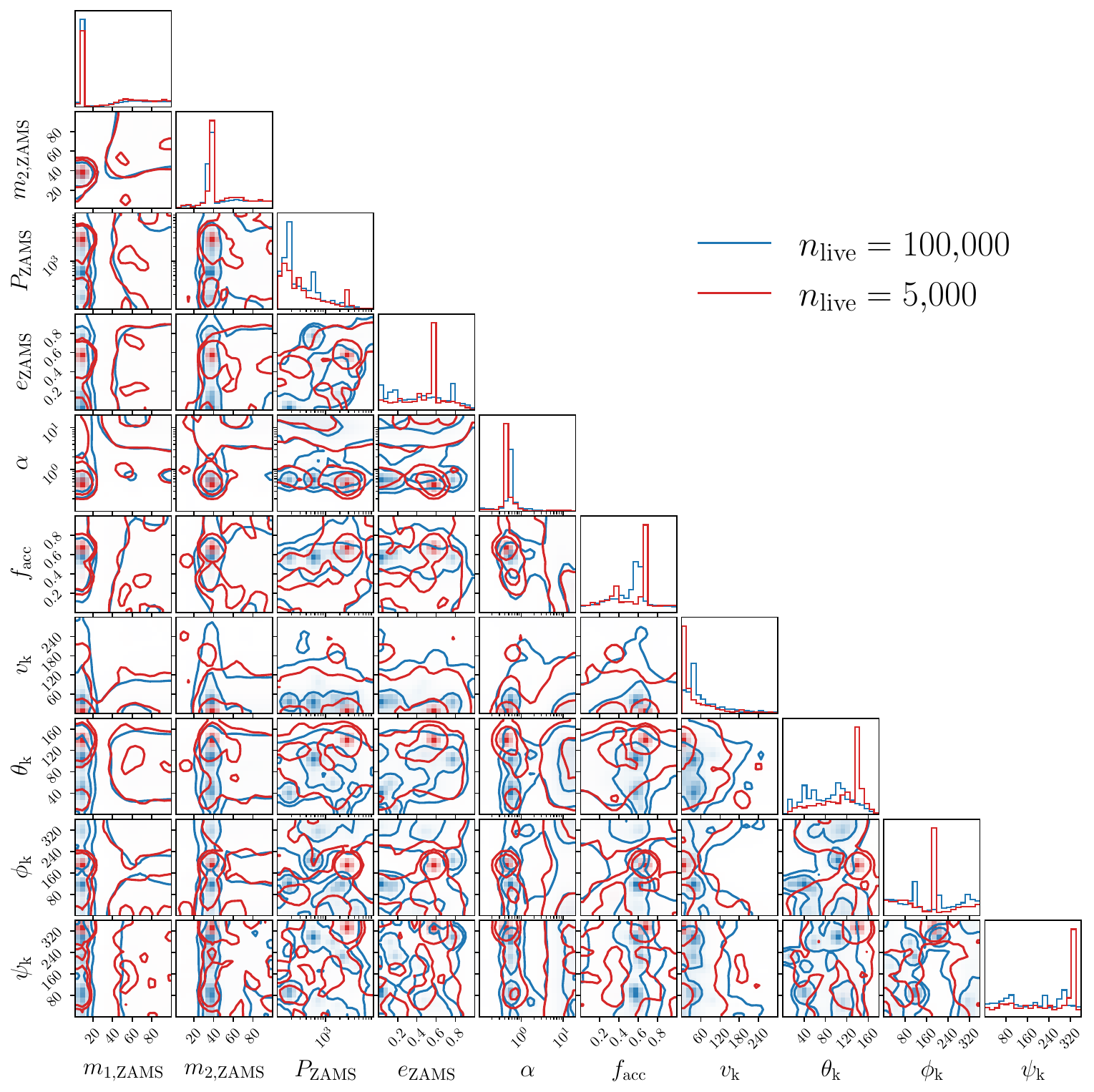}
\caption{Posterior distributions of the ZAMS parameters inferred from our uncalibrated (i.e., without simulation-based calibration) inverse-Bayesian sampling framework: the ZAMS masses, orbital period and eccentricity, the common envelope evolution efficiency $\alpha_{\rm CE}$ and the stable mass transfer accreted fraction $f_{\rm acc}$, and the natal kick velocity magnitude $v_{\rm k}$ and direction in 3-dimensional space. Two datasets are shown corresponding to two numbers of live points, 5,000 in red and 100,000 in blue, with the same fiducial ZAMS prior bounds and all else equal across the two runs that produced these data. The lines in the 2D plots correspond to contours enclosing 50\% and 90\% of the total probability mass.
} \label{F:CornerConverge}
\end{figure*}

\begin{figure*}
\centering
\includegraphics[width=\textwidth]{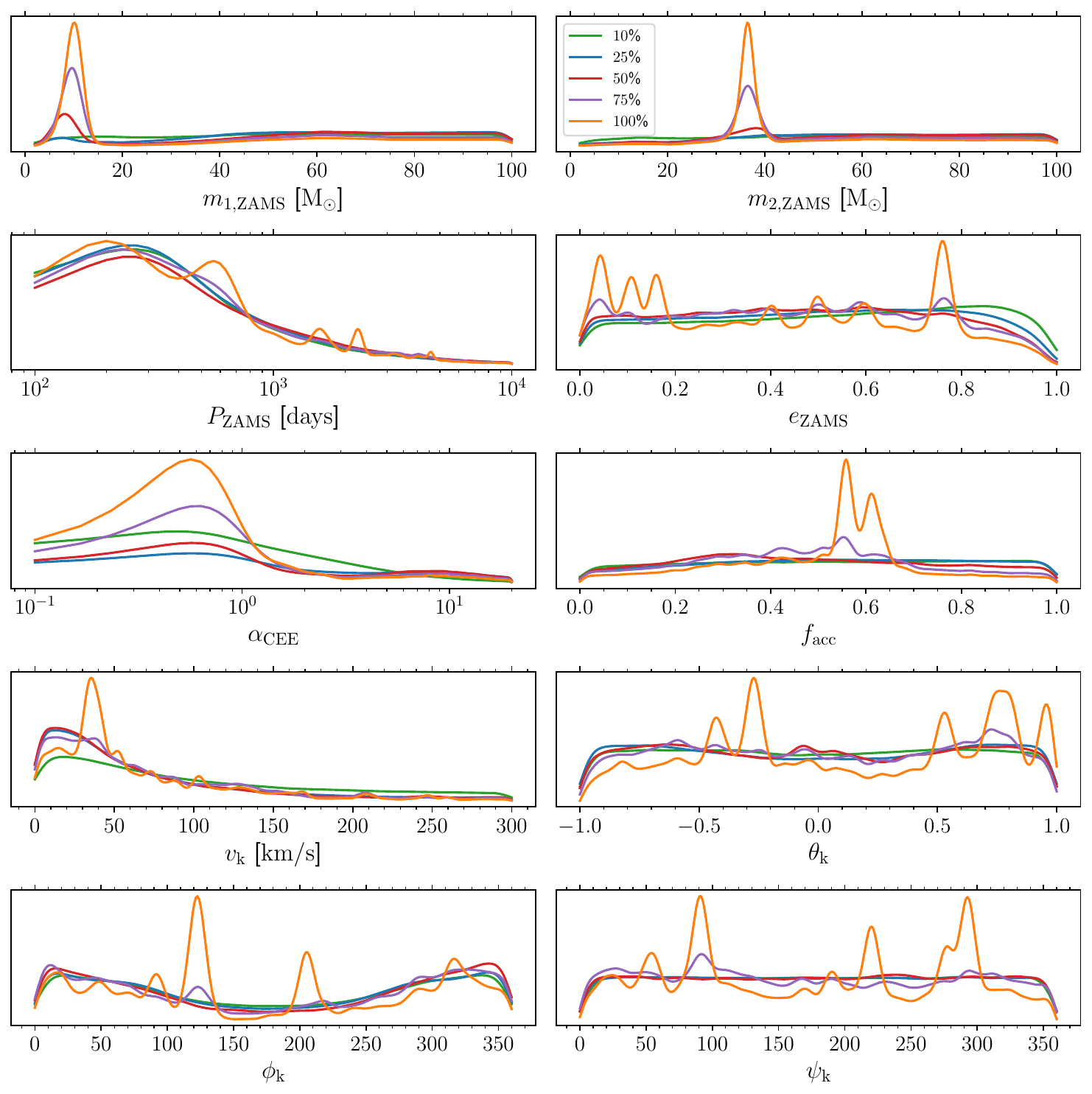}
\caption{Kernel density estimation of subsets of the dataset from Fig.~\ref{F:CornerConverge} with 100,000 live points inferred using fiducial priors, where each line color corresponds to a different fraction (10\%, green; 25\%, blue; 50\%, red; 75\%, purple) of the full dataset (100\%, orange). 
} \label{F:KDE}
\end{figure*}

In this section we present the results of the uncalibrated Bayesian sampler in Section~\ref{Subsec:Uncalibrated} and of the calibrated Bayesian sampler in Section~\ref{Subsec:Calibrated}. 

As a representative example, in Figure~\ref{F:SingleEvo} we first show 
the evolution of a single binary that experiences an SS\,433-like XRB phase where the four panels from top to bottom display the binary masses, orbital period, stellar types, and mass transfer rate. 
This example binary evolves through a canonical pathway of the isolated formation channel of compact binaries in which the primary star ($m_1$; red solid line) loses mass from stellar winds, loses its envelope in common envelope evolution, and forms a black hole (shown by the stellar type of 14 in the third panel). This is followed by a dynamically stable mass transfer event, initiated by the secondary star ($m_2$; blue dashed line), which is modeled as isotropic re-emission where a fraction of the donor star's envelope is accreted and the rest ejected as an isotropic wind. Each of the XRB parameters during this accretion phase pass through the values of SS\,433 in Table~\ref{Tab:Likelihood}

The existence of an XRB phase is most apparent in the masses (top panel) at about $\approx 16.2$ Myr into the binary's lifespan, when the primary star (i.e., the initially more massive star at ZAMS) increases in mass by accreting material from the secondary star during loss of its envelope in a stable mass transfer event, occurs over a short timescale compared to its evolutionary lifetime, and the mass of the primary star correspondingly decreases due to its role as the donor. 
The evolution of the mass transfer rate $M_{\rm dot}$ (bottom panel) during this event shows how this SS\,433-like binary experiences a high amount of mass transfer (green solid line), exceeding the estimated observational mass transfer rate (gray dashed line) of SS\,433 of $10^{-4} \Msol/yr$, for a length of time $\gtrsim 0.01$ Myr. 

The remaining two panels in the middle of Fig.~\ref{F:SingleEvo} show the evolution of the orbital period $P$ and the changing stellar type. 
The common envelope evolution initiated by the primary star is seen as the drastic decrease by a factor of $\approx 1000$ in the period at $\approx 3.6$ Myr, which is followed by a natal kick in BH formation. 
The stellar type ranges over values corresponding to stages of stellar evolution, where each star evolves through the main sequence, possible Hertzsprung gap, and post-main sequence branches, resulting in a core-helium burning star that eventually collapses into a black hole (stellar type = 14) or neutron star (stellar type = 13). 
Combining these with the mass transfer rate implies that this theoretical system satisfies the conditions to be considered SS\,433-like at $\approx 16$ Myr into the evolution of the binary.  

\subsection{\emph{Uncalibrated} progenitor properties of SS~433}
\label{Subsec:Uncalibrated}

As a first-pass analysis, we apply the dynamic nested sampler of \texttt{dynesty} to the likelihood and fiducial ZAMS prior bounds defined in the previous section. We refer to this as an \emph{uncalibrated} analysis because it produces SS\,433 progenitor posterior distributions without SBC iterations, which is the usual approach for Bayesian inference of astrophysical systems \cite{2023ApJ...950..181W,2023MNRAS.518.1057E}. 
As explained in Subsection~\ref{Subsec:Calibrated}, we can parallelize \texttt{dynesty} runs that are without calibration. We implement this via distributed MPI parallelization on the high performance computing cluster Grex. 

We ran a series of parallelized, uncalibrated sampling runs increasing the number of live points of each run. Figure~\ref{F:CornerConverge} shows the resultant inferred ZAMS posterior distributions assuming 5,000 (red) and 100,000 (blue) live points and which contain 135811 and 2710262 many samples, respectively. The convergence of the distributions of ZAMS parameters is clear in the masses, mass transfer parameters, and natal kick velocity magnitude, however the remaining parameters, the period, eccentricity and natal kick angles differ in their dominant modes between the small and large live-point runs. While the blue data with 100,000 live points is able to break degeneracies in the latter set of parameters, the red data with 5,000 live points is unable to; but as we will see in the next section, this can be improved by use of simulation-based calibration. 
The correlation between the orbital period $P_{\rm ZAMS}$ and the natal kick velocity magnitude $v_{\rm k}$ arises from the fact that smaller $v_{\rm k}$ is needed for binaries to survive natal kicks at large $P_{\rm ZAMS}$, as larger orbital periods result in stronger kicks relative to the orbital motion and are hence more likely to unbind binaries. 

Additionally, the parameters that contribute most to the probability of the natal kick of the first-born black hole unbinding the binary---the period, eccentricity and kick parameters---are also the ZAMS parameters most affected by changing the number of live points. 
The distributions of these 5 ZAMS parameters in Fig.~\ref{F:CornerConverge} reveal spiky features corresponding to peaks in the direction of the natal kick that give preferences for binaries that survive their natal kick. 

We investigate these spikes in Figure~\ref{F:KDE}, with the dataset from Fig.~\ref{F:CornerConverge} that used 100,000 live points, where we perform random sampling of the ZAMS posterior distributions and compute kernel-density estimates to obtain four subsets of the full distributions. In each panel of Fig.~\ref{F:KDE}, the orange lines correspond to the full distributions plotted in Fig.~\ref{F:CornerConverge}. The remaining purple, red, blue, and green lines correspond to the subsets where $75\%,\ 50\%,\ 25\%,$ and $10\%$ of the points are sampled, respectively. 
We note the interesting behavior of $\alpha_{\rm CE}$, which is non-monotonic in the fractional size of the series of subsets, i.e., more than $50\%$ of the data points are needed to resolve the main mode at $\alpha_{\rm CE} \approx 0.6$.
This density test verifies that (i) the modes of the distributions of masses and mass transfer parameters are well-converged, and that (ii) the spikes in the orange distributions of the remaining parameters are real features rather than numerical artifacts. To confirm this, we reran the \texttt{dynesty} sampler with 100,000 live points but varying the settings of the sampler itself (i.e., separate runs for different types of walks and boundary conditions and all five of the types of methods implemented in the dynamic nested sampler), but the resultant posterior distributions change negligibly. 
Together, this implies that the spikes are genuine features in the binary evolution parameter space, as determined by \textsc{COSMIC}, indicating that the sampler is capable of breaking degeneracies and locating sharp boundaries due to the complicated assumptions underlying binary evolution prescriptions. 

\newcolumntype{C}[1]{>{\centering\arraybackslash}p{#1}}

\begin{table*}
\caption{
Summary statistics of the simulation-based calibration (SBC) runs for the two cases of initial priors. 
The first case starts with the fiducial priors in Subsection~\ref{Subsec:Optimized}, and the second case with the priors informed from previous studies of SS\,433 (i.e., astro-informed priors) in Subsection~\ref{Subsec:SBC}. 
To help ensure our SBC iterations provide robust tightening of the ZAMS prior bounds while avoiding overfitting, we employ an optimization procedure to search for an optimal number of SS\,433-like binaries $n_{\rm matches}$ after SBC tightening and then to compute new ZAMS prior bounds in each SBC step, balancing the number of available SS\,433-like binaries $n_{\rm selected}$ after resampling, prior volume shrinkage $V_{\rm ratio}$ after tightening, and SS\,433-like purity $p$ (i.e. the fraction of SS\,433-like binaries within the new prior bounds), see Eq.~(\ref{Eq:score}). The procedure also provides an optimized vector of Gaussian widths (one per XRB parameter) used to determine the closeness of inferred binary XRB parameter values to the values of SS\,433 (see Table~\ref{Tab:Likelihood}). This information is used to reliably compute calibrated ZAMS prior bounds which are used for the next run of \textsc{Dynesty}. We repeat this process twice (which effectively produces three sets of refined ZAMS prior bounds) to demonstrate the success of the SBC approach. After an SBC step, we obtain the fraction $f_{\rm in}$ of binaries that are SS\,433-like and within the new calibrated bounds computed at that step. 
}
\label{Tab:SBCresults}
\vspace{0.1cm}
\def\arraystretch{1.2}
\centering
\begin{tabular}{|C{1.5in}||C{0.75in}|C{0.75in}|C{0.85in}|C{0.75in}|C{0.75in}||C{0.75in}|}
  \hline
  SBC step & $n_{\rm matches}$ & $n_{\rm selected}$ & $p$  & $V_{\rm ratio}$ & ${\rm score}$ & $f_{\rm in}$ \\  
  \hline
  fiducial priors & 100,000 & 66,261  &  0.98  & 3.68e-04 & 1.3748 & 0.36 \\  
  iteration 1  & 100,000 & 66,030 & 0.88 & 3.03e-07 & 0.65 & 0.47 \\  
  iteration 2 & 100,000 & 65,338 & 0.76 & 2.52e-07 & 0.55 & 0.83 \\          
  \hline  
  astro-informed priors & 1,000 & 1,000  &  0.03 & 8.4e-14   & -inf  &  0.28 \\    
  iteration 1 & 50,000 & 22,017 & 0.57 & 4.55e-14 & -inf & 0.42 \\     
  iteration 2 & 50,000 & 26,629 & 0.50 & 7.1e-14 & -inf &  0.81 \\       
  \hline
\end{tabular}
\end{table*}

The above results depend on a few assumptions that our exploratory runs reveal as promising avenues for further work: 
\begin{enumerate}
    \item The shapes of the ZAMS posterior distributions are sensitive to the upper bound on the prior of $P_{\rm ZAMS}$. Our fiducial priors have an upper limit of 10,000 days. For larger upper bounds (i.e., 100,000 days), binaries can more easily avoid pre-maturely merging in common envelope evolution, which affects the shape of the distribution of $\alpha_{\rm CE}$ posterior samples, but are less likely to survive large natal kicks in the direction of the orbital motion. 

    \item The shapes of the ZAMS posterior distributions can be sensitive to how many XRB parameters are included in the likelihood definition. For example, we explored runs that did not include $t_{\rm BH}$ in the likelihood, and the sampling generally takes longer compared to the fiducial likelihood that includes it since the extra information helps narrow the space of possible binaries. 

    \item We find that the sampler's ability to find SS\,433-like binaries is sensitive to the assumed  variance of the mass transfer rate of the XRB system, for which we use a conservative variance of $10^{-5} \Msol/yr$. Our simple exploration shows that smaller variance provides for easier sampling (as the sampler looks in a smaller region with higher confidence) up to a limit where it becomes difficult to find any SS\,433-like binaries. Changing the variance by a few orders of magnitude only affect the shapes of the distributions of the ZAMS orbital period, eccentricity, and natal kick angles. 

    \item We find that the sampler's ability to find SS\,433-like binaries is also sensitive to the lower limit on the permissible mass transfer rate in the constraint function which must be specified in order to search for SS\,433-like systems; we assume $10^{-9}\ \Msol/yr$ in all of the results presented herein. A simple exploration with $10^{-6},\ 10^{-7},$ and $10^{-8}\ \Msol/yr$ showed the resultant ZAMS posteriors results did not change significantly but in the case of $10^{-6}\ \Msol/yr$ the sampler struggled to find any acceptable binaries and takes much longer to finish. 

    \item We also explored an analogous problem where we change the \textsc{COSMIC} settings to disallow black hole natal kicks and adjust the sampled ZAMS parameters from the 10 usual ones to the remaining 6 parameters for the masses, orbit, and mass transfer. In these runs, with all else equal to our fiducial, uncalibrated runs, the sampler took an excessively long runtime and did not finish. This suggests that the inclusion of natal kicks provides a harsh discriminator in the binary evolution parameter space, effectively allowing the sampler to more easily eliminate binaries and find a solution. 
\end{enumerate}   
We leave the last two to future work with another sampling approach known as evolutionary or genetic algorithms that can more robustly explore such details in the binary evolution parameter space when Bayesian inference becomes prohibitive \cite{MacMasterSteinle2026inprep}. 

These results motivate the question: how many binaries in our uncalibrated ZAMS posterior distributions are truly SS\,433-like? To answer this quantitatively and robustly, in the following subsection we employ SBC with a pipeline utilizing deterministic and machine learning algorithms (as explained in Subsection~\ref{Subsec:SBC}). 

\begin{figure*}
\centering
\includegraphics[width=\textwidth]{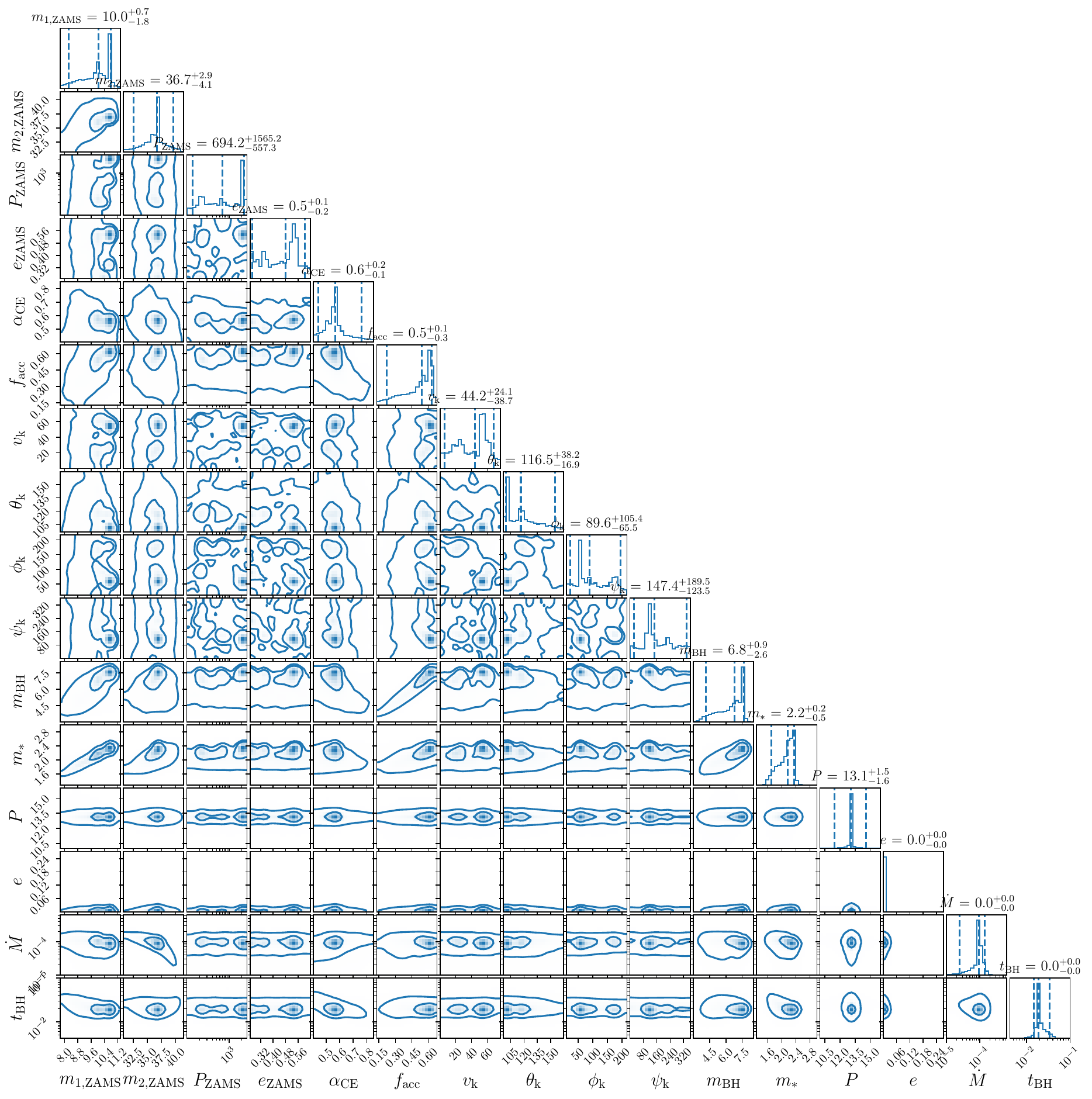}
\caption{Calibrated output after two SBC iterations from the case of fiducial initial priors, i.e., the 83\% of SS\,433-like binaries after the second SBC iteration. The ten ZAMS parameters are shown along with the six XRB parameters. 
} \label{F:CornerCalibrated}
\end{figure*}

\subsection{\emph{Calibrated} progenitor properties of SS\,433}
\label{Subsec:Calibrated}

The results of the previous section utilized MPI parallelization to reduce sampling computation time. But as explained in Sec.~\ref{Subsec:SBC}, our SBC workflow cannot use parallelization due to how the \texttt{dynesty} sampler is constructed, i.e., we must manually and serially save the ZAMS priors and corresponding XRB and ZAMS posteriors. This effectively increases computation time compared to the uncalibrated runs due to appending arrays that are saved at the end of the run. Therefore, to avoid excessive computation times, we limit the SBC runs of this section to a smaller number of live points. Our parameter space exploration in the previous section showed that the uncalibrated \texttt{dynesty} runs are well converged for $\gtrsim 5,000$ live points which we assume in the runs of this section. We consider two series of SBC runs: the first begins with the fiducial ZAMS prior bounds and the second begins with a set of ZAMS prior bounds determined from constraints obtained from other theoretical studies of SS\,433 which we refer to as ``astro-informed'' prior bounds. For both cases we follow the same procedure described in Sec.~\ref{Subsec:SBC}. The results of these runs are summarized in Table~\ref{Tab:SBCresults}. 

In the first case where we begin with the fiducial priors (see Sec.~\ref{Subsec:Optimized}), the optimization procedure selects $n_{\rm matches} = 100,000$ for each iteration and yields satisfactory score. The corresponding optimized $n\sigma$ vector for each iteration has components that vary across the 6 XRB parameters, with the orbital period and mass transfer rate always requiring a $\sigma = 10$ for obtaining SS\,433-like binaries and smaller $\sigma < 10$ for the remaining XRB parameters. 
Each SBC iteration step produces narrower ZAMS prior bounds for some parameters but not all in order to avoid overfitting. 
The near constant purity combined with the decreasing volume ratio $V_{\rm ratio}$ across iterations demonstrate how the SBC procedure guides the sampler to obtain better coverage of SS\,433-like binaries over progressively smaller prior ranges---evidence that the SBC procedure succeeds in calibration. 
The resultant optimized fraction of SS\,433-like binaries, $f_{\rm in}$, correspondingly increases across iterations, implying that the calibration produces more robust SS\,433-like progenitor populations and allows us to build better ZAMS posterior distributions. 
Selecting the 83\% of binaries that are SS\,433-like after the second SBC iteration produces the calibrated ZAMS posterior distributions which we plot in Figure~\ref{F:CornerCalibrated}. These calibrated posterior distributions reflect smoother features in the masses and mass transfer parameters and the spikes in the remaining parameters are also better resolved compared to the uncalibrated result in (the red data of) Fig.~\ref{F:CornerConverge}. 
Although not shown here, we verified that kernel density estimation, analogous to that shown in Fig. 3, of the calibrated ZAMS posteriors confirms the spikes observed in Fig. 4 are true features of the distributions. 
This demonstrates how calibrated \texttt{dynesty} runs are able to break degeneracies in these parameters despite the small number (5,000) of live points used in each SBC step. 

Also included in Fig.~\ref{F:CornerCalibrated} are the distributions of XRB parameters ($m_{\rm BH},\ m_{\rm *},\ P,\ e,\ \dot{\rm M},\ t_{\rm BH}$) whose samples correspond to the forward-evolved binaries of the ZAMS posterior samples. These calibrated XRB parameter samples show tight alignment with the values of SS\,433 used to define the likelihood used in inference (see Table~\ref{Tab:Likelihood}) compared to the uncalibrated (i.e., unmatched) XRB parameter samples from the uncalibrated fiducial run which are far from the values used to define the likelihood (the original motivation for utilizing an SBC procedure). This demonstrates that the SBC procedure is working as intended to produce SS\,433-like systems, with the companion stellar mass as the only exception which produces low-mass XRB systems (with $m_* \approx 2.2 \Msol$) rather than the high-mass XRB of SS\,433. We verified that only 2\% of the uncalibrated ZAMS priors produce high-mass stellar companions (i.e., with $m_* \approx 11.5 \Msol$), indicating that the uncertainty on $m_*$ of $1 \Msol$ used in the likelihood is not sufficient to guide the inference toward high-mass stellar companions, and that the SBC procedure utilized here is limited by the combination of astrophysical assumptions of \textsc{COSMIC} and the fiducial priors. In exploring a preliminary version of an improved calibration procedure (which retains high-weight systems, including rare high-$m_*$ ones), we found that the calibrated ZAMS posterior can develop a second mode at higher donor masses ($\approx 12\Msol$) associated with significantly lower mass-transfer rates ($\dot{M}\sim10^{-7}\Msol\,/{\rm yr}$). Although at first counterintuitive, this behavior is consistent with the binary physics possibilities implemented in \textsc{COSMIC}. For example, high-mass ZAMS donors remain relatively compact for much of their main-sequence lifetime and, when only marginally filling their Roche lobes, can undergo stable, nuclear-timescale mass transfer with comparatively low $\dot{M}$, while lower-mass donors may drive rapid thermal-timescale mass-transfer rates. Consequently, the high-$m_*$, low-$\dot{M}$ mode generically corresponds to ZAMS masses $> 25\Msol$. The appearance of both regimes in the recalibrated posterior reflects the coexistence of two mass-transfer pathways. A more detailed investigation of these possibilities and the prevalence of the high-mass, low-$\dot{M}$ mode is left to future work where we will explore further improvements to the SBC procedure to overcome these limitations and we will compare this analysis with one that uses a different population synthesis model \cite{MacMasterSteinle2026inprep}. 

In the second case, we begin with the following set of astro-informed ZAMS priors with lower bounds of 
$\{25,\ 25,\ 1,\ 0,\ 0.1,\ 0.5,\ 0,\ -1,\ 0,\ 0\}$ and upper bounds of $\{100, 100,\ 1000,\ 0.1,\ 1.0,\ 1.0,\ 100,\ 1,\ 360,\ 360\}$. 
These are derived as follows from \cite{2020ApJ...896...34H} which utilized forward-modeling grid searches with a population synthesis model analogous to \textsc{COSMIC}. 
The ZAMS eccentricity and orbital period and the mass transfer parameters are limited to within the ranges listed above, as \cite{2020ApJ...896...34H} assume circular ZAMS orbits (see their Sec. 2.1), ZAMS semi-major axis between 3 and 1000 ${\rm R}_\odot$ (corresponding to $P_{\rm ZAMS} \approx 0.01$ and 585 days, respectively), conservative accreted fraction $f_{\rm acc}$ (where their model and \textsc{COSMIC} employ the isotropic re-emission mode of SMT \cite{2014LRR....17....3P}) of 0.5 to 1.0 (see their Sec. 2.2, accretion efficiency modes II and III), and modest CEE efficiency $\alpha_{\rm CE}$ of 0.1 to 1.0 (see their Sec. 2.1). 
Also, from the analysis of \cite{2024ApJ...975L..28C}, we use $m_{\rm ZAMS} > 25 \Msol$ (based on supernova explosion and remant simulations) which excludes low-mass ZAMS progenitors. 
We were unable to find refinement for the natal kick parameters of SS\,433 in the literature, but the small measured eccentricity (i.e. see Table~\ref{Tab:Measurements}) suggests a not-too strong natal kick and hence we assume 3-dimensional kicks with $v_{\rm k} < 100$ km/s. 

The astro-informed SBC runs are more challenging to obtain improvements in sampling SS\,433-like binaries compared to the case of fiducial priors due to the narrower initial prior space. 
This is shown in the failure to find an optimal $n_{\rm matches}$ (indicated by the score of -inf) and corresponding $n\sigma$ vector. 
Instead we manually inspect the tightening output and choose a conservative estimate to be able to carry out the SBC, focusing on purity rather than parameter space coverage since the prior volume is already small. We still obtain increased $f_{\rm in}$ after each iteration, allowing us to build better ZAMS posterior distributions of SS\,433-like binaries. While this case is more challenging to build smooth posterior distributions compared with SBC on the fiducial prior bounds, we obtain ZAMS posterior distributions that display similar features. 

We also repeated these analyses and checked with a train/validation split of the datasets, instead of using the full datasets for training (i.e. calibration) and validation as in Table~\ref{Tab:SBCresults}, but this showed no change in the summary statistics and in the calibrated ZAMS prior bounds. This was further verified by running a series of split-data training/validation checks, where the difference between the training and validation purities is $\approx 0$. More checks and improvements can be performed in this framework, which we leave to future work. 

Lastly, as a simple comparison, we also employ \texttt{emcee} \cite{2013PASP..125..306F} for MCMC sampling instead of \texttt{dynesty} nested sampling, as our modular code allows for simple exchanging of such samplers for the same likelihood function and constraints. 
This shows that the \texttt{emcee} sampler produces markedly different output distributions compared to the \texttt{dynesty} sampler. On the surface, these \texttt{emcee} ZAMS distributions look more smooth as they do not display spikes and they display peaks at different values of the ZAMS parameters. However, \texttt{emcee} generically fails to produce reliable SS\,433-like binaries, which we verified by running the SBC analysis on the uncalibrated \texttt{emcee} output and find that, even with the most optimistic settings, the purity is always $\approx 0$. This implies that one must utilize caution when applying bare samplers to the highly complex, degenerate, and multi-dimensional parameter spaces of binary evolution as certain samplers, like MCMC samplers, are prone to finding solutions far from the target used to define the likelihood, limiting reproducibility of results of studies that only use bare MCMC sampling. However, we did not try advanced MCMC techniques, such as parallel-tempered MCMC, which might improve its performance. 

\section{Conclusions}
\label{sec:Conclusions}

We implement a Bayesian approach to infer the progenitor distribution of SS\,433, in which we sample over ZAMS binary parameters conditioned on the observed properties of the system. While not Bayesian optimization in the strict surrogate-model sense, our use of likelihood-based nested sampling represents a Bayesian optimization–inspired strategy that focuses computational effort on high-likelihood regions of progenitor space.

Our approach advances a growing body of work applying Bayesian and machine-learning frameworks to constrain the progenitor populations of compact-object binaries. These methods differ in their use of surrogate likelihoods or machine-learned approximations of the forward model. By contrast, our work uses a fully forward-modeling approach with no surrogate or emulator, and a likelihood constructed directly from the observed parameters of the X-ray binary SS\,433. We apply nested sampling over a ten-dimensional space of ZAMS binary parameters, using the rapid binary population synthesis code \textsc{COSMIC} as the forward mapping. Our methodology provides direct posterior samples over progenitor space, offering interpretable predictions and avoiding approximation error from learned models. 
However, we found that such approaches can suffer in their predictions relative to the target properties of the observed source. To combat this, we employed a new statistical technique known as simulation-based calibration (SBC) to iteratively build robust posterior distributions of the progenitors of SS\,433-like systems. Our current SBC procedure will be improved in upcoming work \cite{MacMasterSteinle2026inprep} to overcome challenges identified here, most important of which is ensuring accurate recovery of the mass of the stellar companion. 

Our main conclusions are: 
\begin{itemize}
    \item Inverse modeling via Bayesian inference is a powerful tool for astrophysics, but must be paired with other methods such as SBC for robust predictions. In our uncalibrated runs, an extremely large number (100,000) of live points was needed to probe degeneracies in the period, eccentricity, and natal kick direction; in our SBC runs, which are unparallelizable due to the need of serially saving all ZAMS priors and ZAMS and XRB posteriors outside of the \texttt{dynesty} evaluations, we show that calibration can still break such degeneracies with a smaller---but still large---number (5,000) of live points. 
    
    \item We find that the BH progenitor ZAMS mass is $< 25 \Msol$ for binaries with a low donor mass $m_*$ and high mass transfer rate. However, an improved SBC procedure shows that the BH progenitor ZAMS mass is $> 25 \Msol$, in agreement with the result of \cite{2024ApJ...975L..28C}, for binaries with a high donor mass $m_*$ and low mass transfer rate. This results in a trade-off where binaries with high donor mass $m_*$, and hence producing HMXRBs, have a mass transfer rate that is less than the value that defines an SS\,433-like binary. The bimodal nature of our calibrated posteriors, as detailed in Section III.B., is an important and intriguing outcome of the SBC approach. We also attempted sampling runs assuming that BHs do not receive natal kicks, which proved challenging despite the lower dimensional parameter space. We leave both of these details for further investigation in upcoming future work that will employ evolutionary/genetic algorithms and improved SBC \cite{MacMasterSteinle2026inprep}. 
    
    \item The sampler's ability to find SS\,433-like binaries is sensitive to the assumed variance of the mass transfer rate in the likelihood Eq.~(\ref{Eq:SamplePosterior}), which corresponds to the measurement uncertainty on the observed mass transfer rate of SS\,433 (see Table~\ref{Tab:Measurements}). This motivates future observational campaigns to provide improved constraints on the mass transfer rate of the binary in SS\,433. For example, if the future space-based AXIS telescope \citep{2023SPIE12678E..1ER} can constrain the uncertainty by at least one or two orders of magnitude smaller than the current measurement it would provide a potential performance boost for a model such as ours, a topic we leave to upcoming work \cite{MacMasterSteinle2026inprep}. 
\end{itemize}
We note that our choice of the \textsc{COSMIC} population synthesis model stems from its modular design which allows for a high amount of binary specification in the likelihood evaluation, and the use of other models may lead to different results, although many such models are based on the same or similar assumptions. 

In future work, we plan to incorporate jet properties, which will offer valuable information for our progenitor inference framework. 
SS\,433 and its nebula W50 constitute the most robust Galactic example of jet-driven particle acceleration up to PeV energies 
\citep{SafiHarb2022ApJ935, MacIntyreetal2025inprep, 2025arXiv251006431T}.
The observed jet morphology, spectral hardness gradients, and environmental feedback encode key information about the kinetic energy output, age, and ambient density. These parameters likely correlate with progenitor characteristics, 
suggesting a promising path forward for expanding the analysis presented in this study to include environmental information, such as jet–ISM interaction, similar to the analysis in \cite{2020ApJ...896...34H}. 
In addition to spectral and morphological properties, geometrical aspects of the jets of SS\,433 could be useful, for example by including the jet precession period of 
$\approx 162$ days \citep{2021MNRAS.507L..19C} and the half-opening angle of $\approx$20 deg \citep{2004ApJ...616L.159B}. These observables are likely governed by combinations of the binary mass ratio, orbital separation, and accretion disk tilt, with the precession period and jet cone angle potentially set by tidal or relativistic coupling between the disk and the binary orbit, implying the need for a model that relates the binary progenitor evolution to the jet convolved with the progenitor population model. 
With such a model, instead of modifying the likelihood function directly, one could constrain the population synthesis output in a similar way as done in our study in the likelihood evaluation. However, such a task is nontrivial, given the large uncertainties in jet launching mechanisms and their relation to kinematically-driven jet evolution and to population properties of stellar binaries. 

We note that the recent discovery by LHAASO of ultra-high-energy (UHE; \(E>100\) TeV) $\gamma$-ray emission from multiple Galactic micro-quasar systems, including SS\,433, V4641 Sgr, GRS 1915+105, MAXI J1820+070, and Cygnus X-1, has transformed our understanding of microquasars as extreme particle accelerators \citep{2024arXiv241008988L}. In particular, the detection of spatially‐extended UHE emission from SS\,433 in coincidence with a massive gas cloud strongly favors a hadronic origin and suggests that microquasar jets can accelerate protons (or heavier nuclei) up to PeV energies, making them viable Galactic PeVatrons \citep{2024arXiv241008988L}. The independent observations of V4641 Sgr by LHAASO and H.E.S.S. and the detection of extended X-ray emission around it by XRISM further support the notion that such systems may not be exceedingly rare \cite{2024Natur.634..557A,2025arXiv251110537A,2025ApJ...978L..20S}. 
    
Taken together with theoretical modeling of the Galactic micro-quasar population \cite{2025arXiv251001369K}, these developments suggest that only a handful ($\sim$10) of active, powerful micro-quasars in the Galaxy at any given time could suffice to explain a substantial fraction of the cosmic-ray proton flux near the ``knee.'' While our present work focuses on the progenitor and formation constraints for SS\,433-like systems, emerging evidence underscores that estimating the Galactic population of such systems is critical to assess their overall contribution to high-energy cosmic rays. 

A fully quantitative estimate of the number of SS\,433–like systems in the Galaxy would involve combining our inferred formation constraints with a Galactic-binary population model and a realistic Galactic potential. As a first step toward this goal one could integrate orbits of synthetic binaries through a Milky Way potential, accounting for natal kicks and spatial evolution. 
Beyond the scope of this work, we leave this promising and timely question ---combining inferred birthrates and binary-evolution predictions with the observed PeVatron census--- to a future study.

\acknowledgements
We thank the anonymous referee for their exceptionally insightful comments and feedback, which helped to strengthen the impact of our study. 
N.S., S.S.H., A.M., and B.M.I. are supported by the Natural Sciences and Engineering Research Council of Canada (NSERC) through the Canada Research Chairs and Discovery Grants programs. B.M.I. further acknowledges support from the NSERC CGS-D fellowship. M.M. is supported by the Royal Commission for the Exhibition of 1851 Research Fellowship. 
Computations described in this paper were performed using the \href{https://umanitoba.ca/information-services-technology/research-computing/um-high-performance-computing-system-grex}{University of Manitoba Grex High Performance Computing Centre} (RRID:SCR\_026342), which provides an HPC service to the University’s research community. 

\bibliography{SS433}
\end{document}